\documentclass[twocolumn,showpacs,aps,prl,superscriptaddress]{revtex4}

\usepackage{graphicx}
\usepackage{dcolumn}
\usepackage{amsmath}
\usepackage{epsfig}
\usepackage[dvips]{feynmp}

\RequirePackage{xspace}

\usepackage{relsize}
\def\babar{\mbox{\slshape B\kern-0.1em{\smaller A}\kern-0.1em
    B\kern-0.1em{\smaller A\kern-0.2em R}}}

\def\t     {\ensuremath{t}\xspace}

\def\piz   {\ensuremath{\pi^0}\xspace}

\def\pip   {\ensuremath{\pi^+}\xspace}

\def\Kbar  {\kern 0.2em\overline{\kern -0.2em K}{}\xspace}

\def\Kz    {\ensuremath{K^0}\xspace}
\def\Kzb   {\ensuremath{\Kbar^0}\xspace}
\def\KzKzb {\ensuremath{\Kz \kern -0.16em \Kzb}\xspace}
\def\Kp    {\ensuremath{K^+}\xspace}
\def\Km    {\ensuremath{K^-}\xspace}

\def\KpKm  {\ensuremath{\Kp \kern -0.16em \Km}\xspace}
\def\KS    {\ensuremath{K^0_{\scriptscriptstyle S}}\xspace}

\def\Dbar    {\kern 0.2em\overline{\kern -0.2em D}{}\xspace}

\def\Dz      {\ensuremath{D^0}\xspace}
\def\Dzb     {\ensuremath{\Dbar^0}\xspace}
\def\DzDzb   {\ensuremath{\Dz {\kern -0.16em \Dzb}}\xspace}
\def\Dp      {\ensuremath{D^+}\xspace}
\def\Dm      {\ensuremath{D^-}\xspace}

\def\Dmp     {\ensuremath{D^\mp}\xspace}
\def\DpDm    {\ensuremath{\Dp {\kern -0.16em \Dm}}\xspace}
\def\Dstar   {\ensuremath{D^*}\xspace}

\def\Dstarp  {\ensuremath{D^{*+}}\xspace}

\def\Dstarpm {\ensuremath{D^{*\pm}}\xspace}

\def\B       {\ensuremath{B}\xspace}
\def\Bbar    {\kern 0.18em\overline{\kern -0.18em B}{}\xspace}

\def\BB      {\ensuremath{B\Bbar}\xspace} 
\def\Bz      {\ensuremath{B^0}\xspace}
\def\Bzb     {\ensuremath{\Bbar^0}\xspace}
\def\BzBzb   {\ensuremath{\Bz {\kern -0.16em \Bzb}}\xspace}
\def\Bu      {\ensuremath{B^+}\xspace}
\def\Bub     {\ensuremath{B^-}\xspace}

\def\BpBm    {\ensuremath{\Bu {\kern -0.16em \Bub}}\xspace}

\def\jpsi     {\ensuremath{{J\mskip -3mu/\mskip -2mu\psi\mskip 2mu}}\xspace}

\mathchardef\Upsilon="7107
\def\Y#1S{\ensuremath{\Upsilon{(#1S)}}\xspace}

\mathchardef\Deltares="7101
\mathchardef\Xi="7104
\mathchardef\Lambda="7103
\mathchardef\Sigma="7106
\mathchardef\Omega="710A

\def\Deltabar{\kern 0.25em\overline{\kern -0.25em \Deltares}{}\xspace}
\def\Lbar{\kern 0.2em\overline{\kern -0.2em\Lambda\kern 0.05em}\kern-0.05em{}\xspace}
\def\Sigbar{\kern 0.2em\overline{\kern -0.2em \Sigma}{}\xspace}
\def\Xibar{\kern 0.2em\overline{\kern -0.2em \Xi}{}\xspace}
\def\Obar{\kern 0.2em\overline{\kern -0.2em \Omega}{}\xspace}
\def\Nbar{\kern 0.2em\overline{\kern -0.2em N}{}\xspace}
\def\Xb{\kern 0.2em\overline{\kern -0.2em X}{}\xspace}

\def\bpsiks     {\ensuremath{\Bz \to \jpsi \KS}\xspace}

\def\mes        {\mbox{$m_{\rm ES}$}\xspace}

\newcommand{\tev}{\ensuremath{\mathrm{\,Te\kern -0.1em V}}\xspace}
\newcommand{\gev}{\ensuremath{\mathrm{\,Ge\kern -0.1em V}}\xspace}
\newcommand{\mev}{\ensuremath{\mathrm{\,Me\kern -0.1em V}}\xspace}
\newcommand{\kev}{\ensuremath{\mathrm{\,ke\kern -0.1em V}}\xspace}
\newcommand{\ev}{\ensuremath{\mathrm{\,e\kern -0.1em V}}\xspace}
\newcommand{\gevc}{\ensuremath{{\mathrm{\,Ge\kern -0.1em V\!/}c}}\xspace}
\newcommand{\mevc}{\ensuremath{{\mathrm{\,Me\kern -0.1em V\!/}c}}\xspace}
\newcommand{\gevcc}{\ensuremath{{\mathrm{\,Ge\kern -0.1em V\!/}c^2}}\xspace}
\newcommand{\mevcc}{\ensuremath{{\mathrm{\,Me\kern -0.1em V\!/}c^2}}\xspace}

\def\mus  {\ensuremath{\rm \,\mus}\xspace}

\def\mus        {\ensuremath{\,\mu{\rm s}}\xspace}    

\def\to                 {\ensuremath{\rightarrow}\xspace}

\def\pep2{PEP-II}

\def\gsim{{~\raise.15em\hbox{$>$}\kern-.85em
          \lower.35em\hbox{$\sim$}~}\xspace}
\def\lsim{{~\raise.15em\hbox{$<$}\kern-.85em
          \lower.35em\hbox{$\sim$}~}\xspace}

\def\CP                {\ensuremath{C\!P}\xspace}

\def\stwob{\ensuremath{\sin\! 2 \beta   }\xspace}

\def\deltaz{\ensuremath{{\rm \Delta}z}\xspace}
\def\deltat{\ensuremath{{\rm \Delta}t}\xspace}
\def\deltamd{\ensuremath{{\rm \Delta}m_d}\xspace}

\xspace

\newcommand{\jprBase}        {Phys.\ Rev.\xspace}

\newcommand{\nimBaseC}       {Nucl.\ Instr.\ and Methods\xspace}

\newcommand{\npBase}         {Nucl.\ Phys.\xspace}
\newcommand{\zpBase}         {Z.\ Phys.\xspace}

\newcommand{\nim}       [1]  {\nimBaseC~{\bf #1}}
\newcommand{\np}        [1]  {\npBase\ {\bf #1}}
\newcommand{\pr}        [1]  {\jprBase\ {\bf #1}}

\newcommand{\progtp}    [1]  {{Prog.\ Th.\ Phys.\ {\bf #1}}}

\newcommand{\zp}        [1]  {\zpBase\ {\bf #1}}

\def\jetset74   {\mbox{\tt Jetset \hspace{-0.5em}7.\hspace{-0.2em}4}\xspace}

\def\Dstarp     {\ensuremath{D^{*+}}\xspace}

\def\Dstarpm    {\ensuremath{D^{*\pm}}\xspace}

\def\Dp         {\ensuremath{D^+}\xspace}
\def\Dm         {\ensuremath{D^-}\xspace}

\def\Bztodstdcc {\ensuremath{B^{0} \to D^{*\pm} D^{\mp}}\xspace}

\def\DeltaEStd  {\ensuremath{\Delta E} \xspace}

\newcommand{\BABARPubYear}    {03}
\newcommand{\BABARPubNumber}  {004}

\newcommand{\SLACPubNumber} {9661}
\newcommand{\LANLNumber} {0303004}

\def\figurebox#1#2#3{%
    \def\arg{#3}%
    \ifx\arg\empty
    {\hfill\vbox{\hsize#2\hrule\hbox to #2{\vrule\hfill\vbox to #1{\hsize#2\vfill}\vrule}\hrule}\hfill}%
    \else
    {\hfill\epsfbox{#3}\hfill}%
    \fi}

\begin{document}

\preprint{\babar-PUB-\BABARPubYear/\BABARPubNumber} 
\preprint{SLAC-PUB-\SLACPubNumber} 

\begin{flushleft}
\babar-PUB-\BABARPubYear/\BABARPubNumber\\
SLAC-PUB-\SLACPubNumber\\
hep-ex/\LANLNumber\\[10mm]
\end{flushleft}

\title{
{\Large \bf \boldmath
Measurement of the Branching Fraction and \CP-violating Asymmetries in 
Neutral $B$ Decays to $D^{*\pm}D^{\mp}$
}}

%
\author{B.~Aubert}
\author{R.~Barate}
\author{D.~Boutigny}
\author{J.-M.~Gaillard}
\author{A.~Hicheur}
\author{Y.~Karyotakis}
\author{J.~P.~Lees}
\author{P.~Robbe}
\author{V.~Tisserand}
\author{A.~Zghiche}
\affiliation{Laboratoire de Physique des Particules, F-74941 Annecy-le-Vieux, France }
% ...
\collaboration{The \babar\ Collaboration}
\noaffiliation

\date{\today}

\begin{abstract}
We present measurements of the branching fraction and \CP-violating asymmetries for
neutral $B$ decays to $D^{*\pm}D^{\mp}$.
The measurement uses a data sample of approximately 88 million \Y4S $\to$ \BB decays
collected with the \babar\ detector at the SLAC PEP-II asymmetric-energy $e^{+}$-$e^{-}$ collider.  By fully reconstructing the
$D^{*\pm}D^{\mp}$ decay products, we measure the
branching fraction to be $(8.8 \pm 1.0 \pm 1.3) \times 10^{-4}$
and the time-integrated \CP-violating asymmetry between the rates to $D^{*-}D^{+}$ and $D^{*+}D^{-}$ to be
$\mathcal{A} = -0.03 \pm 0.11 \pm 0.05$.  We also measure the
time-dependent \CP-violating asymmetry parameters to be $S_{\scriptscriptstyle -+} = -0.24 \pm 0.69 \pm 0.12$, 
$C_{\scriptscriptstyle -+} = -0.22 \pm  0.37 \pm 0.10$ for $B \to D^{*-}D^{+}$ and $S_{\scriptscriptstyle +-} = -0.82 \pm 0.75 \pm 0.14$, 
$C_{\scriptscriptstyle +-} = -0.47 \pm 0.40 \pm 0.12$ for $B \to D^{*+}D^{-}$.
In each case the first error is statistical and the second error is systematic.
\end{abstract}


\maketitle

Within the Standard Model (SM) of electroweak interactions, \CP violation is the result of a complex phase 
in $V$, the Cabbibo-Kobayashi-Maskawa (CKM) quark mixing matrix~\cite{CKM}.  
In the SM, the time-dependent \CP-violating asymmetries in 
$B \to D^{*\pm}D^{\mp}$
decays are related to the angle
$\beta \equiv \arg[-V_{\rm cd}V^{*}_{\rm cb}/V_{\rm td}V^{*}_{\rm tb}]$.  
We
present a measurement of the branching fraction and a first measurement of \CP-violating asymmetries in
$B \to D^{*\pm}D^{\mp}$ decays using a sample of $87.9 \pm 1.0$ million \BB decays.

As recent measurements of the parameter \stwob using the quark process $b \to c\bar{c}s$ have shown, \CP is
violated in the neutral $B$-meson system~\cite{BabarSin2b,BelleSin2b}.  The measured asymmetries are currently
consistent with the SM 
expectation~\cite{CKMWorkshop}.
In order to search for additional sources of \CP violation from new physics processes, different quark decays such as 
$b\to c\bar{c}d$ must be examined.

In $b \to
c\bar{c}d$ processes (for example, 
$B \to D^{*\pm}D^{\mp}$ decays; see Fig.~1), 
penguin contributions containing a different weak phase than the tree
are not expected to be as highly suppressed
as in $b \to c\bar{c}s$ decays; thus the relation of the time-dependent 
\CP-violating asymmetries in $b \to c\bar{c}d$ decays to $\beta$ is less exact than in decays such as \bpsiks.  
However, the contribution from additional weak phases in time-dependent asymmetries in $b \to c\bar{c}d$ due 
to purely SM processes is still expected to be fairly small, of order $\Delta \beta = 0.1$ in a simplified 
model~\cite{GrossWorah,Gronau}.  
A variety of beyond-SM processes, which can provide additional sources of \CP violation, can greatly 
increase this contribution, up to $\Delta \beta \approx 0.6$ in some models~\cite{GrossWorah}.

\vspace*{-0.6cm}
\setlength{\unitlength}{.31cm}
\begin{figure}[h]
\hspace*{-0.7cm}
\begin{minipage}[h]{3.0cm}
\includegraphics{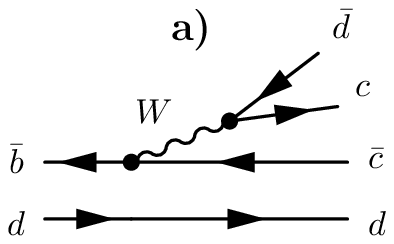}
\end{minipage}
\hspace*{1.0cm}
\begin{minipage}[h]{3.0cm}
\vspace*{0.5cm}
\includegraphics{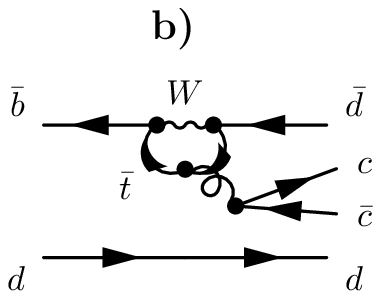}
\end{minipage}
\vspace*{-0.0cm}
\caption{The leading-order Feynman graphs for \Bztodstdcc decay: 
a) tree diagram and b) penguin diagram.}
\end{figure}


\CP-violating asymmetries in 
$B \to D^{*\pm}D^{\mp}$
are due to interference between the decay amplitudes and the \BzBzb
mixing amplitude, as well as interference between tree and penguin decay amplitudes. 
The decay rate distributions $f^{\pm}$, where
the superscript $+(-)$ refers to whether the tagging meson, $B_{\rm tag}$, was \Bz(\Bzb), are given by
\begin{eqnarray}
f^{\pm} & = & \frac{e^{-\left|\deltat\right|/\tau}}{4\tau} \times \nonumber \\
 & & [1 \pm S \sin(\deltamd\deltat) \mp C \cos(\deltamd\deltat)],
\label{fplusminus}
\end{eqnarray}
where $\tau$ is the mean \Bz lifetime, $\deltamd$ is the $\Bz\Bzb$ mixing frequency,
and $\deltat = t_{\rm reco} - \t_{\rm tag}$ is the time elapsed between the 
$B$ decay to 
$D^{*\pm}D^{\mp}$ and the decay of $B_{\rm tag}$.  Separate $S$ and $C$ parameters
are fitted for the two decays $D^{*-}D^{+}$ and $D^{*+}D^{-}$, resulting in the four fitted \CP-violation parameters
$\{ S_{\scriptscriptstyle -+}, C_{\scriptscriptstyle -+}, S_{\scriptscriptstyle +-}, C_{\scriptscriptstyle +-} \}$.  The 
time-integrated asymmetry $\mathcal{A}$ between the rates to $D^{*-}D^{+}$ and $D^{*+}D^{-}$
is defined as
\begin{equation}
\mathcal{A} = \frac{N_{D^{*+}D^{-}} - N_{D^{*-}D^{+}}}{N_{D^{*+}D^{-}} + N_{D^{*-}D^{+}}}.
\end{equation}

The states $D^{*-}D^{+}$ and $D^{*+}D^{-}$ are not 
\CP eigenstates.  The formalism of time evolution for non-\CP eigenstate vector-pseudoscalar decays 
is given in Ref.~\cite{Aleksan}.  
In the case of equal amplitudes for $B \to D^{*-}D^{+}$ and $B \to D^{*+}D^{-}$, one expects that at tree level
$C_{\scriptscriptstyle -+} = C_{\scriptscriptstyle +-} = 0$ and $S_{\scriptscriptstyle -+} =
S_{\scriptscriptstyle +-} = -\stwob$.


A detailed description of the \babar\ detector is presented in Ref.~\cite{BABARNIM}.  Charged particle 
momenta are measured in a tracking system consisting of a 5-layer double-sided silicon vertex tracker (SVT) 
and a 40-layer hexagonal-cell wire drift chamber (DCH) filled with a gas mixture of helium and isobutane.  The 
SVT and DCH operate within a 1.5 T solenoidal field.  
Photons 
are detected and their energies are measured in a CsI(Tl) electromagnetic calorimeter.  Muons 
are identified in 
the instrumented flux return (IFR), composed of resistive plate chambers and layers of iron that 
return the magnetic flux of the solenoid.  A detector of internally reflected Cherenkov light (DIRC) 
is used for particle identification.

We select hadronic events using 
track multiplicity and event topology criteria.  At least three 
reconstructed 
tracks, each with transverse momentum greater than 100 MeV/$c$, are required in the 
laboratory polar angle region $0.41 < \theta_{\rm lab} < 2.54$, where $\theta_{\rm lab} = 0$ is the $e^{-}$-beam direction.  
The event must have a total measured energy in the laboratory frame greater than 4.5 GeV.  In order 
to help reject non-\BB background, the ratio of Fox-Wolfram moments $H_2/H_0$ is required to be less than 0.5~\cite{fox}.

For reconstruction of 
$B \to D^{*\pm}D^{\mp}$ decays,
all daughter tracks are required to pass within 10 cm in $z$ and 1.5 cm in 
$r-\phi$ of the center of the beam crossing region.
A track is identified as a charged kaon candidate using the 
Cherenkov angle measured in the DIRC and energy loss information (d$E$/d$x$) from the DCH and SVT.

Neutral pion candidates are composed of pairs of photons in the EMC.  The photons must each have energy above 30 MeV, 
and their energy must sum to greater than 200 MeV.
The \piz candidates must have an invariant mass between 115 and 150 MeV/$c^2$.  A mass-constrained fit is 
imposed on \piz candidates, 
in order to improve resolution on the energy of reconstructed $B$ candidates.

We require \KS $\to \pi^{+}\pi^{-}$ candidates to have an invariant mass within 15 MeV/$c^2$ of the nominal \KS
mass~\cite{PDG2002}.  The transverse flight distance of the \KS from the primary event vertex is required to be greater than
2 mm.

To form $D$ candidates, 
kaon candidates are combined with other tracks, assumed to be pions, and \piz candidates in the event.  
We reconstruct \Dz candidates in the four modes
$K^{-}\pi^{+}$, $K^{-}\pi^{+}\piz$, $K^{-}\pi^{+}\pi^{-}\pi^{+}$, and $\KS\pi^{+}\pi^{-}$, and \Dp candidates
in the modes $K^{-}\pi^{+}\pi^{+}$ and $\KS\pi^{+}$.
We require \Dz and \Dp candidates to have reconstructed invariant masses within
20 MeV/$c^2$ of their respective nominal masses, except for \Dz decays with a \piz daughter,
which must be within 35 MeV/$c^2$ of the nominal \Dz mass.  
Mass-constrained fits are applied to \Dz and \Dp candidates in order to improve the measurement of the
momentum of each $D$.  The \Dstarp is then reconstructed in its decay to $\Dz\pip$.

To select neutral $B$ candidates from pairs of well-reconstructed
\Dstarpm and \Dmp candidates, we form a likelihood that includes all measured \Dstar and $D$ masses:
\begin{eqnarray}
\mathcal{L} & = & G(m_{\Dmp},\sigma_{m_{\Dmp}}) \cdot G(m_{\Dz},\sigma_{m_{\Dz}}) \cdot \nonumber \\
            &   &  H(\delta m_{\Dstarpm},\sigma^{\rm core}_{\delta m_{\Dstarpm}},\sigma^{\rm tail}_{\delta m_{\Dstarpm}}, 
                     f_{\rm core}),
\end{eqnarray}
where the \Dstarpm - \Dz mass difference is denoted by $\delta m_{\Dstar}$.
Each $G$ represents a Gaussian distribution, and $H$ is the sum of two Gaussian distributions, for the core and tail of 
the 
$\delta m_{\Dstarpm}$ distribution, respectively.
For $\sigma_{m_D}$ we use
values individually computed for each $D$ candidate, while for $\sigma_{\delta m_{D^{*}}}$
we use values obtained from an inclusive \Dstar data sample: 0.35\mevcc for the core Gaussian distribution and 1.27\mevcc
for the tail, and a core fraction ($f_{\rm core}$) of 51\%.  
Likelihood cuts are set individually for each 
combination of \Dstarpm and \Dmp decay modes,
using a detailed Monte Carlo simulation, in order to maximize the expected 
signal sensitivity.
In events with 
more than one \Bz candidate, we choose the candidate with the highest likelihood value.

A 
$B \to D^{*\pm}D^{\mp}$
candidate is characterized by two kinematic variables:
the beam-energy substituted mass,
$\mes \equiv \sqrt{(\sqrt{s}/2)^{2} - {p_B^*}^2}$,
and the difference of the $B$ candidate's measured energy from the beam energy, 
$\DeltaEStd \equiv E_{B}^* - (\sqrt{s}/2)$.
$E_{B}^*$ ($p_B^*$) is the energy (momentum) of the \B\ candidate
in the $e^{+}e^{-}$ center-of-mass frame and $\sqrt{s}$ is the
total center-of-mass energy.
The signal region in \DeltaEStd is defined to be $|\DeltaEStd| < 18\mev$. 
According to Monte Carlo simulations,
the width of this region corresponds to approximately 
twice the signal resolution
in \DeltaEStd.  

The 
$B \to D^{*\pm}D^{\mp}$
decay candidates in the region $5.27 < \mes < 5.30 \gevcc$ and 
$|\DeltaEStd| < 18\mev$ are used to extract \CP-violating asymmetries.
A sideband, defined as $5.20 < \mes < 5.27 \gevcc$ and $|\DeltaEStd| 
< 18\mev$, and a ``large sideband,'' defined as $5.20 < \mes < 5.27 \gevcc$ and 
$|\DeltaEStd| < 200\mev$, are used to extract various background parameters.  The total
numbers of selected events in the signal region, the sideband, and the large sideband 
are 197, 461, and 5187, respectively.

To extract the number of signal events above background, as well as the time-integrated 
\CP asymmetry $\mathcal{A}$ 
(see Eq.~2),
we use an unbinned extended
maximum likelihood fit to the \mes distribution of the $D^{*\pm}D^{\mp}$ candidates, including the sideband.
%
%
The \mes distribution for the simultaneous fit to
all the selected events is described by Gaussian distributions for the $D^{*+}D^{-}$ and $D^{*-}D^{+}$
signals, an ARGUS threshold function~\cite{Argus}, and a Gaussian distribution to describe a small
potential ``peaking'' background contribution (due to $B$ decays such as \Bz $\to D^{*-}D_s^{+}$ that
are similar to the signal modes).  The endpoint of the ARGUS function is fixed to the average
beam energy.  From studies performed with both data and Monte Carlo simulations, the ``peaking''
contribution is estimated to be $12 \pm 8$ events.
There are a total of four free parameters in the nominal fit: the shape and normalization of the background ARGUS 
function (2), the total 
$B \to D^{*\pm}D^{\mp}$ signal yield (1), and the \CP asymmetry $\mathcal{A}$ (1).
The total 
$B \to D^{*\pm}D^{\mp}$ signal yield is determined to be $113 \pm 13$ events. 
Figure 2 shows the \mes distributions for
$B \to D^{*-}D^{+}$ and $D^{*+}D^{-}$ candidates.

We use a Monte Carlo simulation of the \babar\ detector to
determine the efficiency for reconstructing the 
$B \to D^{*\pm}D^{\mp}$ signal.  The
efficiencies range from 6\% to 18\%, depending on the $D$ decay   
modes.  From these efficiencies and the total number of recorded \BB pairs, 
and assuming the $\Y4S \to \BzBzb$ branching fraction to be $50\%$, we determine 
the branching fraction for 
neutral $B$ to $D^{*\pm}D^{\mp}$ to be
$$\mathcal{B}(B \to D^{*\pm}D^{\mp}) = (8.8 \pm 1.0{\rm (stat.)} \pm 1.3{\rm (syst.)}) \times 10^{-4}.$$

\begin{figure}[!htb]
\begin{center}
\includegraphics[width=8.5cm]{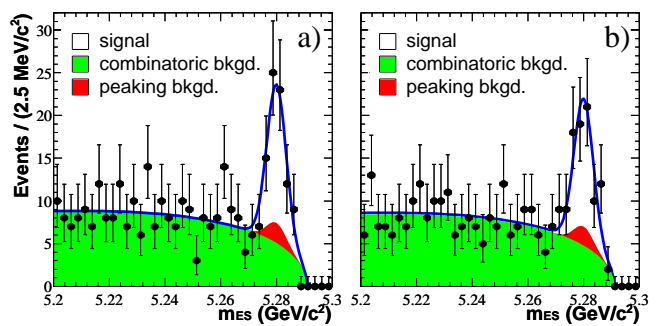}
\caption{
The \mes distributions of a)
$B \to D^{*-}D^{+}$
and b) $B \to D^{*+}D^{-}$
candidates with $|\DeltaEStd| < 18\,\mev$.
The fit includes Gaussian distributions to model the signal and 
a small peaking background component, and an ARGUS
function~\cite{Argus} to model the combinatoric background shape.
}
\label{fig:mes}   
\end{center}
\end{figure}

Systematic uncertainties on the branching fraction are dominated by uncertainty on the charged-particle
tracking efficiency ($8.9\%$), uncertainties on the branching fractions of the $D$ decay modes ($7.4\%$)~\cite{PDG2002},
and the uncertainty on the amount of peaking background ($6.8\%$).  The total 
systematic uncertainty from all considered sources is $14.5\%$.  The result is consistent with Ref.~\cite{BelleDstD}.

The fitted value for $\mathcal{A}$ is 
$$\mathcal{A} = -0.03 \pm 0.11{\rm (stat.)} \pm 0.05{\rm (syst.)}.$$

Systematic uncertainties on $\mathcal{A}$ are dominated by potential differences in the reconstruction efficiencies 
of positively and negatively charged tracks (0.04), and by uncertainty in the \mes resolution for
$B \to D^{*\pm}D^{\mp}$ signal events (0.03).

The method for extracting time-dependent \CP asymmetries shares many of the techniques that are used for the 
measurement of \stwob in
charmonium decays in \babar~\cite{BabarSin2b}.  We use the same algorithms for determination of the flavor of the 
tagging $B$ in the event, for determining the distance \deltaz between the 
$B \to D^{*\pm}D^{\mp}$
and tagging $B$ decay
vertices, and for performing the maximum likelihood fit.  We also use the same data sample, $B_{\rm flav}$, of
fully reconstructed $B$ decays to $D^{(*)\pm}(\pi^{\mp}, \rho^{\mp}, a_1^{\mp})$ to measure tagging performance and
\deltaz resolution.

The $B$ flavor-tagging algorithm relies on the correlation between the flavor of the $b$ quark and the particle types, momenta,
and charges of the remaining tracks in the event.  A multivariate algorithm is used to separate events into four
tagging categories and to determine tag flavor, the details of which are given in Ref.~\cite{BabarSin2b}.  

The elapsed time \deltat between the 
$B \to D^{*\pm}D^{\mp}$
and tagging $B$ decays is determined from the measured distance \deltaz between the $z$ positions of
the two $B$ decay vertices and from the known boost of the $e^+e^-$ system.  A detailed description of the algorithm is given in 
Ref.~\cite{BabarS2bPRD}.  We accept events with $\sigma_{\deltat} <$ 2.5\hspace{.08cm}ps and 
$|\deltat| <$ 20\hspace{.08cm}ps, where 
$\sigma_{\deltat}$ is the error on \deltat.  We find that 93\% of signal candidates satisfy these requirements.

We determine the time-dependent \CP asymmetry parameters 
using a simultaneous unbinned maximum likelihood fit to the \deltat distributions of the
$D^{*\pm}D^{\mp}$ and $B_{\rm flav}$ candidates,
including \mes sideband samples for background parametrization.  The \deltat distribution for $D^{*\pm}D^{\mp}$ signal 
events is described by
Eq.~1.  The \deltat distribution of $B_{\rm flav}$ events 
is also described by Eq.~1 with $C=1$ and $S=0$ where the superscript $+(-)$ refers to opposite (same) flavor events, comparing
the reconstructed and tag $B$ mesons.
The mistag fraction $w$ reduces the measured 
$S$ and $C$ coefficients
by a factor $1 - 2w$; this fraction is measured 
within the fit for each tagging category, utilizing the large $B_{\rm flav}$ sample.  We convolve the \deltat
distribution with a resolution function modelled by the sum of three Gaussian distributions.  The \deltat resolution is 
dominated by the
tag vertex $z$-position resolution and is parametrized 
in the same way as
for the charmonium \stwob measurement; this is described in detail in Ref.~\cite{BabarS2bPRD}.  Both
continuum and \BB backgrounds are incorporated, each with a \deltat distribution that is determined
within the fit, using the \mes sideband.

There are 37 fitted parameters in the combined fit for time-dependent \CP asymmetries: the \CP asymmetry parameters 
$S_{\scriptscriptstyle -+}$, $C_{\scriptscriptstyle -+}$, $S_{\scriptscriptstyle +-}$, and $C_{\scriptscriptstyle +-}$ (4); 
the average mistag fractions $w_i$ (4), and the differences 
$\Delta w_i$ between \Bz and \Bzb mistag fractions (4), where $i$ is one of the four tagging categories; parameters for 
the signal \deltat resolution 
function (8); and parameters for background time dependence (6), \deltat resolution (3), and mistag fractions (8).  
The $B_{\rm flav}$ sample constrains all parameters except the \CP asymmetries.  In the nominal fit, we fix 
$\tau_{\Bz} =$ 1.542\hspace{.08cm}ps and $\Delta m_d =$ 0.489\hspace{.08cm}ps$^{-1}$~\cite{PDG2002}.  

The time-dependent \CP asymmetry fit to the 
$B \to D^{*\pm}D^{\mp}$
and $B_{\rm flav}$ samples yields
\begin{eqnarray}
S_{\scriptscriptstyle -+} & = &  -0.24 \pm 0.69{\rm (stat.)} \pm 0.12{\rm (syst.)}, \nonumber\\
C_{\scriptscriptstyle -+} & = &  -0.22 \pm 0.37{\rm (stat.)} \pm 0.10{\rm (syst.)}, \nonumber\\
S_{\scriptscriptstyle +-} & = &  -0.82 \pm 0.75{\rm (stat.)} \pm 0.14{\rm (syst.)}, \nonumber\\
C_{\scriptscriptstyle +-} & = &  -0.47 \pm 0.40{\rm (stat.)} \pm 0.12{\rm (syst.)}. \nonumber
\end{eqnarray}
The correlation between $S_{\scriptscriptstyle -+}$ and $C_{\scriptscriptstyle -+}$ is $0.16$ and between $S_{\scriptscriptstyle +-}$ and 
$C_{\scriptscriptstyle +-}$ is $-0.01$.  Besides these correlations, the magnitudes of all correlations of the $S$ and $C$ parameters with any 
other free parameter are each less than 0.04.
Figure 3 shows the \deltat distributions and asymmetries in yields between \Bz and \Bzb tags
for the $D^{*-}D^{+}$ and $D^{*+}D^{-}$ samples,
each overlaid with a projection of the fit result.

\begin{figure}[!htb]
\begin{center}
\includegraphics[width=7.5cm]{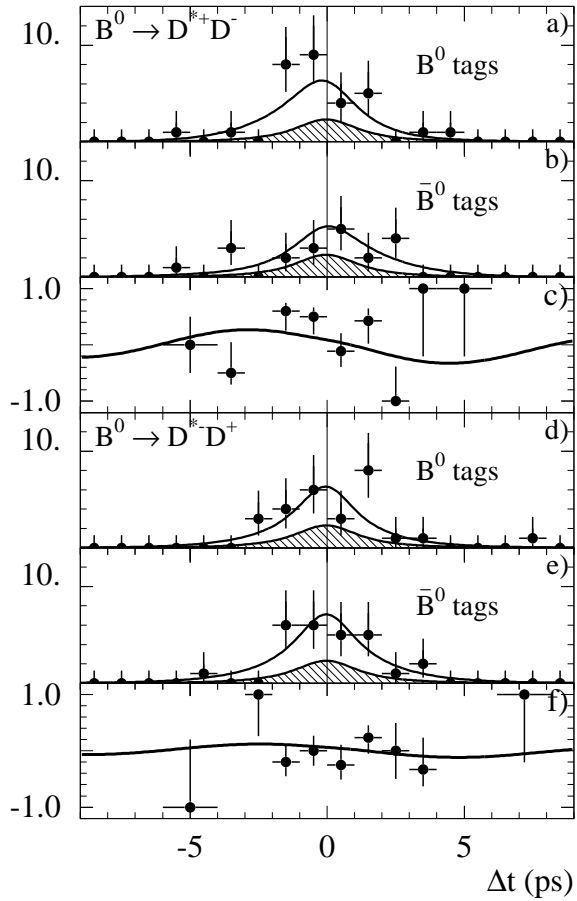}
\caption{
Distributions of \deltat for 
$B \to D^{*+}D^{-}$
candidates in the signal region with a) a \Bz tag ($N_{\Bz}$) and b) with a \Bzb tag
($N_{\Bzb}$), and c) the raw asymmetry $(N_{\Bz} - N_{\Bzb})/(N_{\Bz} + N_{\Bzb})$.  The
solid curves are the fit projections in \deltat.  The shaded regions represent the
background contributions.  Figures d), e), and f) contain the corresponding information for 
$D^{*-}D^{+}$.
}
\label{fig:dtasym}
\end{center}
\end{figure}

Systematic uncertainties on the time-dependent \CP asymmetry parameters are dominated by uncertainties in the amount,
composition, and \CP asymmetry of the background in the selected $D^{*\pm}D^{\mp}$ events (resulting in errors on the 
parameters ranging from 0.07-0.10); the parametrization of the
\deltat resolution function (0.01-0.06); possible differences between the $B_{\rm flav}$ and $D^{*\pm}D^{\mp}$ mistag fractions (0.01-0.04);
the error on a small correction to the fitted asymmetries due to 
the limited size
of the $D^{*\pm}D^{\mp}$ sample (0.01-0.02); and the potential presence of a small amount of \CP-violating interference between leading 
order and 
doubly-CKM-suppressed decay channels of the tagging $B$ meson (0.01-0.03).  

In summary, we have measured 
the branching fraction and \CP-violating asymmetries for
$B \to D^{*\pm}D^{\mp}$
decays.  The small size of the $D^{*\pm}D^{\mp}$ sample currently precludes the observation of \CP violation 
in this first measurement in this 
channel; however, with the addition of more data, future results may  
provide important information about sources of \CP-violation in the $B$-meson system.

We are grateful for the excellent luminosity and machine conditions
provided by our \pep2\ colleagues, 
and for the substantial dedicated effort from
the computing organizations that support \babar.
The collaborating institutions wish to thank 
SLAC for its support and kind hospitality. 
This work is supported by
DOE
and NSF (USA),
NSERC (Canada),
IHEP (China),
CEA and
CNRS-IN2P3
(France),
BMBF and DFG
(Germany),
INFN (Italy),
FOM (The Netherlands),
NFR (Norway),
MIST (Russia), and
PPARC (United Kingdom). 
Individuals have received support from the 
A.~P.~Sloan Foundation, 
Research Corporation,
and Alexander von Humboldt Foundation.

\end{document}